\documentclass[namedreferences]{SolarPhysics}
\usepackage[optionalrh]{spr-sola-addons} 
\usepackage{graphicx}        
\usepackage{color}           
\usepackage{url}             
\usepackage[english]{babel}
\usepackage[breaklinks=true]{hyperref}

\newcommand{\aap}{    {\it Astron. Astrophys.}}
\newcommand{\aaps}{   {\it Astron. Astrophys. Suppl.}}

\newcommand{\apj}{    {\it Astrophys. J.}}
\newcommand{\apjl}{    {\it Astrophys. J. Lett.}}

\newcommand{\solphys}{{\it Solar Phys.}}

\begin{document}

\begin{article}

\begin{opening}

\title{Oscillations in  the Solar Faculae. III. The Phase Relations between Chromospheric and
Photospheric LOS Velocities}

\author{N.I.\,\surname{Kobanov}$^{1}$\sep
	A.S.\,\surname{Kustov}$^{1}$\sep
        S.A.\,\surname{Chupin}$^{1}$\sep
        V.A.\,\surname{Pulyaev}$^{1}$
       }
\runningauthor{N.I. Kobanov \textit{et al}.}
\runningtitle{Oscillations in Solar Faculae. III}

   \institute{$^{1}$ Institute of Solar-Terrestrial Physics, Irkutsk, P.O.\,Box\,291,  Russia\\ (email: \url{kobanov@iszf.irk.ru)}\\
             }

\begin{abstract} An analysis of line-of-sight velocity oscillation in nine solar faculae was undertaken with
the aim of studying of phase relations between chromosphere (He\,{\sc i} 10\,830\,\AA{} line) and photosphere (Si\,{\sc i} 10\,827\,\AA{} line)
five-minute oscillations. We found that time lag of the chromospheric signal relative to photospheric one varies from -12 to 100 seconds
and is about 50 seconds on average.We assume that the small observed lag can have three possible explanations: \textit{i})~convergence of formation levels of He\,{\sc i} 10\,830\,\AA{} and Si\,{\sc i} 10\,827\,\AA{} in faculae; \textit{ii})~significant increase of five-minute oscillation propagation velocity above faculae; \textit{iii})~simultaneous presence of standing and travelling waves.
\end{abstract}
\end{opening}

\section{Introduction}
     \label{S-Introduction}

Solar faculae, usually called \textquotedblleft plages\textquotedblright in the photosphere or calcium and hydrogen flocculi in the chromosphere,
are often observed in the solar atmosphere. They may play a significant role in energy exchange processes
 between layers of the solar atmosphere due to their prevalence. One of possible ways to transport energy is waves
propagating in faculae. The facular oscillations have been actively studied since the 1960s \cite{Orrall65,Howard67,Bhatnagar71,Deubner74,Teske74,Woods80}.

{\inlinecite{Orrall65} found no evidence of a difference between the velocity amplitudes underlying weak plages or the chromospheric network, and regions free of calcium flocculi. There was only a slight indication that the periods are a little longer in photospheric regions that underlie the network. \inlinecite{Howard67} found that the amplitudes of five-minute oscillations are about
25\% weaker in regions with a significant magnetic field ($>80$\,G). \inlinecite{Teske74} compared power spectra of velocities in plages with velocity power spectra of quiet photosphere and suggested that photospheric oscillations are not gravity waves. \inlinecite{Balthasar90} found that the power of five-minute range oscillations can both increase and decrease in faculae.The analysis of line-of-sight (LOS) velocity oscillations in  Fe\,{\sc i} 15\,648\,\AA{} and 15\,652\,\AA{} spectral lines \cite{Muglach95} revealed  dominant oscillatory signal in the velocity data is due to five-minute range oscillations. Also temporal variations of the magnetic-field strength were observed which exhibit a possible five-minute and nine\,--\,ten-minute oscillation \cite{Muglach95}.}

{\inlinecite{Khomenko08} and \inlinecite{Centeno09} found that the time delay between photospheric (Si\,{\sc i} 10\,827\,\AA{}) and chromospheric (He\,{\sc i} 10\,830\,\AA{}) five-minute oscillations in faculae is 300\,--\,500 seconds. In their opinion, this is a evidence of linear vertical propagation of acoustic waves.
\inlinecite{Kobanov11} noticed that  the five-minute LOS-velocity oscillations observed in H$\alpha$ and Fe\,{\sc i} 6569\,\AA{} show  ambiguous phase lags in faculae.}

This article is the third in a series dealing with the investigation of features of oscillations in faculae.
It was preceded by \inlinecite{Kobanov07} and \inlinecite{Kobanov10}, hereafter Paper I and Paper II, respectively.
In this article we focus on the determination of time lag between signals
of the LOS velocity measured in the He\,{\sc i} 10\,830\,\AA{} and Si\,{\sc i} 10\,827\,\AA{} spectral lines.

Direct measurements of the time lag between chromospheric and photospheric signals of the LOS velocity are of
interest to test the assumption that oscillations propagate upwards.

\section{Instruments and Methods} 
      \label{S-Instruments}

The observational spectral data analyzed here were obtained with the horizontal solar telescope of the Sayan Solar Observatory.
The photoelectric guider of the telescope is capable of tracking the solar image with an accuracy of 1 arcsecond for
several hours of observations and compensates for the rotation of the Sun \cite{Kobanov09}.
We used a Princeton Instruments RTE/CCD 256H camera 256$\times$1024 pixels with a pixel size of 24 microns.
One pixel corresponds to 0.24 arcsecond along the entrance slit of the spectrograph and about 12 m\AA{} along the dispersion.
Thus, the spectral snapshot contains information about a spatial region of about 60$''$$\times$1$''$.
The real resolution of the telescope is 1\,--\,1.5$''$ because of the Earth's atmosphere.  Seeing and stray light were assessed according to the contrast of granulation and the limb in the visible. The intensity of stray light in the infrared range at the altitude of the observatory (2000 m) is low. For 10\,830\,\AA{} its intensity is roughly 16 times less than for 5000\,\AA{}.
We made observations in the He\,{\sc i} 10\,830\,\AA{} and Si\,{\sc i} 10\,827\,\AA{} spectral lines.

The LOS velocity in the He\,{\sc i} and Si\,{\sc i} lines was measured by the method that is known as
the \textquotedblleft Doppler compensation method\textquotedblright \cite{Severny58}. A pair of virtual slits were placed symmetrically on the spectrogram on both sides of
the line centre ($\pm$0.2\,\AA{} for He\,{\sc i} and $\pm$0.18\,\AA{} for Si\,{\sc i}) in such a way that the corresponding intensities $\mathrm{\textit{I}_{blue}}$
and $\mathrm{\textit{I}_{red}}$ could be the same.  Intensity in He\,{\sc i} was calculated as $\mathrm{\textit{I}=(\textit{I}_{red} +\textit{I}_{blue})/2\textit{I}_c} $, where $\mathrm{\textit{I}_c}$ -- continuum intensity. When the spectral line was shifted, the virtual slits were displaced in the same
direction, equalizing the intensities. The displacement  is proportional to the Doppler velocity. The initial position
of the virtual slits (zero point) was determined relative to the telluric line $\mathrm{H_2O}$  10\,832\,\AA{}, that was used to
eliminate the spectrograph noise.

With a decrease in distance between the virtual slits (up to $\pm$0.1\,\AA{}), no significant changes in He\,{\sc i} signal were noted. Only insignificant changes in signal/noise ratio, associated with a decrease in the steepness of the working part of the line profile, were noted. For Si\,{\sc i} 10\,827\,\AA{} the picture was similar. For the both spectral lines behaviour of the LOS-velocity signal  did not change appreciably. The width of virtual slits was 0.06\,\AA{} for both lines.

\section{Observations}
       \label{S-Observations}

\begin{figure}
  \centerline{\includegraphics[clip=]{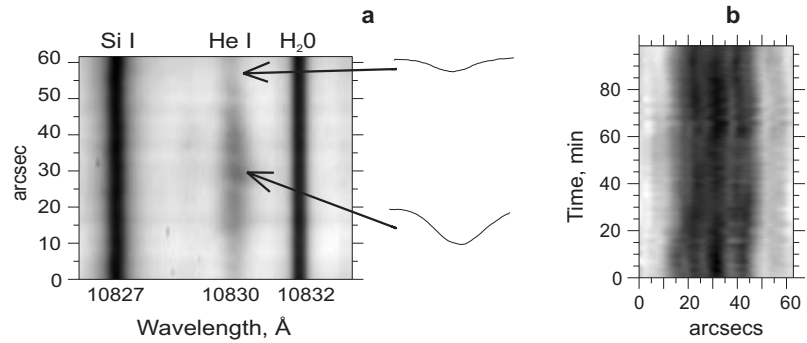}
              }
\caption{Increasing depth He\,{\sc i} 10\,830\,\AA{} in solar faculae (facula 5). a)Fragment of spectrogram and profiles of He\,{\sc i} 10\,830\,\AA{}
in different image parts; b)grey-scale spatial--time diagram of He\,{\sc i} intensity.}
  \label{F-FaculaFeature}
\end{figure}

Compared to undisturbed regions, the depth of the He\,{\sc i} 10\,830\,\AA{} line profile in faculae sharply increases (Figure~\ref{F-FaculaFeature}a).
According to our observations, this increase in different faculae varies from 2 to 5.
This feature of the He\,{\sc i} line is very useful for precise pointing at faculae near the disk centre.
At the beginning, we chose an object with the use of images in the Ca\,{\sc ii}\,H line and determined its coordinates.
Then we slowly scanned the chosen region and corrected its position in the spectrograph slit, using the working
camera and taking the maximum depth of the He\,{\sc i} 10\,830\,\AA{} line into consideration. We also should take into account
the fact that the  depth of this line increases in filaments. That is why we used  H$\alpha$ filtergrams
for additional control. In most cases, the facular region was rather inhomogeneous and occupied the major part of the
entrance aperture. Compact faculae were observed more seldom. In the present analysis we have used faculae in which the residual intensity in the core of He\,{\sc i} 10\,830\,\AA{} did not exceed 0.8 for the space of 20$''$.

It is noteworthy that space--time diagrams of the He\,{\sc i} 10\,830\,\AA{}  intensity can be used to
control position of the observed object in the spectrograph slit  throughout a time series (Figure~\ref{F-FaculaFeature}b).

In the Summer 2010, we obtained 33 time series for 24 faculae near the disk centre; this allowed us to
avoid the influence of projection effects when comparing photospheric and chromospheric signals. The most complete and multiple-factor analysis given in the article was carried out on basis of nine time series. We selected those series that corresponded to better seeing and minimal blurring. In addition, we took into account the following criteria: the remoteness from sunspot no less than 60$''$, the facular size no less than 20$''$, and homogeneity of facula (the residual intensity in the core of He\,{\sc i} 10\,830\,\AA{} no more than 0.8), as well as the duration of the series no less than 50 minutes.

Preprocessing of spectrograms consisted of standard procedures: removal of the \textquotedblleft dust\textquotedblright effect and determination of
the flat field. Using special programmes, we then plotted space--time diagrams of the He\,{\sc i} 10\,830\,\AA{} intensity and
diagrams showing the spatial distribution of different-frequency modes in the 1\,--\,8 mHz range for photospheric and
chromospheric signals of the LOS velocity. With the use of these diagrams, we determined facular fragments in
which five-minute oscillations are present both in the photosphere and chromosphere. We detected five-minute oscillations in the  2.5\,--\,4.5 mHz band. The significance level for the power of LOS-velocity oscillations in this band was 95\%. The LOS-velocity signals in
the facular fragments of 2$''$ located at a distance of 5$''$ from boundaries were then subjected to frequency filtering in the 1 mHz band centred at 3.5 mHz.
Our chosen parameters of frequency filtering corresponded to those of \inlinecite{Centeno09} in order to compare
results properly. The average amplitude of the photospheric signal was then increased up to the average amplitude
of the chromospheric signal, and the average lag of the chromospheric signal relative to the photospheric one was
determined using cross-correlation. Table\,\ref{T-TimeSeries} presents obtained lag values and the amplitude multiplication factor of the
photospheric LOS-velocity signal $\mathrm{\textit{K}=\textit{A}_{ch}/\textit{A}_{ph}}$ for each time series. The coefficient $\textit{K}$ relates only to the LOS velocities filtered in the 3\,--\,4 mHz band.

\begin{table}
\begin{tabular}{cccccccc}
\hline  \\ {\bf No.}&{\bf Date}&{\bf Time,}&{\bf Disk}&{\bf Cadence,}&{\bf Duration,}&{\bf \slshape K}&{\bf Lag $\mathbf{[\delta t]}$,}\\
{\bf}&{\bf}&{\bf UT }&{\bf location}&{\bf sec}&{\bf min }&{\bf}&{\bf sec} \\
\hline
\\ 1 & 17 Jul 10 & 01:40 & 19$^\circ$N  28$^\circ$W & 3 & 150 & 6& 78\,--\,100
\\ 2 & 04 Aug 10 & 09:48 & 15$^\circ$N  05$^\circ$W & 4 & 81 & 1.2& -12\,--\,0
\\ 3 & 05 Aug 10 & 00:28 & 20$^\circ$N  14$^\circ$W & 3.5 & 96 & 1.5& 20\,--\,40
\\ 4 & 09 Aug 10 & 00:45 & 17$^\circ$N  23$^\circ$E & 3 & 136 &1.8& 40\,--\,80
\\ 5 & 09 Aug 10 & 06:10 & 18$^\circ$N  22$^\circ$E & 4 & 98 & 2.9& 10\,--\,50
\\ 6 & 09 Aug 10 & 09:49 & 15$^\circ$N  12$^\circ$E & 3 & 68 & 1.7& 68\,--\,77
\\ 7 & 14 Aug 10 & 07:29 & 13$^\circ$N  05$^\circ$W & 3 & 198 &2.2& 30\,--\,90
\\ 8 & 15 Aug 10 & 04:24 & 24$^\circ$N  18$^\circ$E & 3 & 102&3.3
& 33\,--\,39
\\ 9 & 16 Aug 10 & 01:58 & 32$^\circ$N 00$^\circ$E & 3 & 55&8
& 68\,--\,78
\\ \hline

\end{tabular}
\caption[]{Information on observational data, results of measurement of time lag and $K$.}
\label{T-TimeSeries}
\end{table}

\section{Analysis Results. Discussion}
       \label{S-Discussion}

Our previous articles (Paper I, Paper II, and \opencite{Kobanov11}) dealt mainly with observations in the H$\alpha$ 6563\,\AA{} and Fe\,{\sc i} 6569\,\AA{} spectral lines.
According to \inlinecite{Vernazza81}, the height of the H$\alpha$ line formation is 1550\,--\,2000 km, whereas that of the
Fe\,{\sc i} 6569\,\AA{} line is 150\,--\,250 km \cite{Parnell69}. The difference in formation heights of these lines is about
1500 km. It is noteworthy that, according to \inlinecite{Centeno09}, the difference in heights in faculae for the He\,{\sc i} 10\,830\,\AA{}
and Si\,{\sc i} 10\,827\,\AA{} lines is 1500 km. Spectra of chromospheric and photospheric oscillations of LOS velocity in faculae in \inlinecite{Khomenko08}
and \inlinecite{Centeno09} coincide down to the smallest details in the 1\,--\,8 mHz range.
The time lag of the chromospheric signal relative to the photospheric one at 3.5 mHz is 300\,--\,400 seconds. Our results for
the H$\alpha$ and Fe\,{\sc i} 6569\,\AA{} lines \cite{Kobanov11} contradict those of Centeno and Khomenko. Photospheric and chromospheric spectra differ
 significantly in the low-frequency (1\,--\,2 mHz) and high-frequency (5\,--\,7 mHz) regions. The power of chromospheric oscillations
at 1\,--\,2 mHz frequencies obtained in  H$\alpha$ is very often comparable or exceeding the power of five-minute oscillations. Similarity between spectra
is observed only in the central part of the range (2.5\,--\,4.5 mHz).

But the main problem is that five-minute oscillations observed by us in the H$\alpha$ and Fe\,{\sc i} 6569\,\AA{}
lines do not show time lags of the chromospheric signal of 300\,--\,400 seconds, unlike  \inlinecite{Centeno09}.
In the majority of the phase-difference measurements, the chromospheric signal lags behind the photospheric signal;
the lag value is however small and is rarely as large as 100 seconds. In some facular fragments, the chromospheric signal at some
time intervals was slightly in advance relative to the photospheric signal. This should be nominally interpreted
as downward propagation \cite{Kobanov11}.

Explanation for these contradictions is hampered by the fact that different pairs of spectral lines were
used in the observations.

\begin{figure}
 \centerline{\includegraphics[clip=]{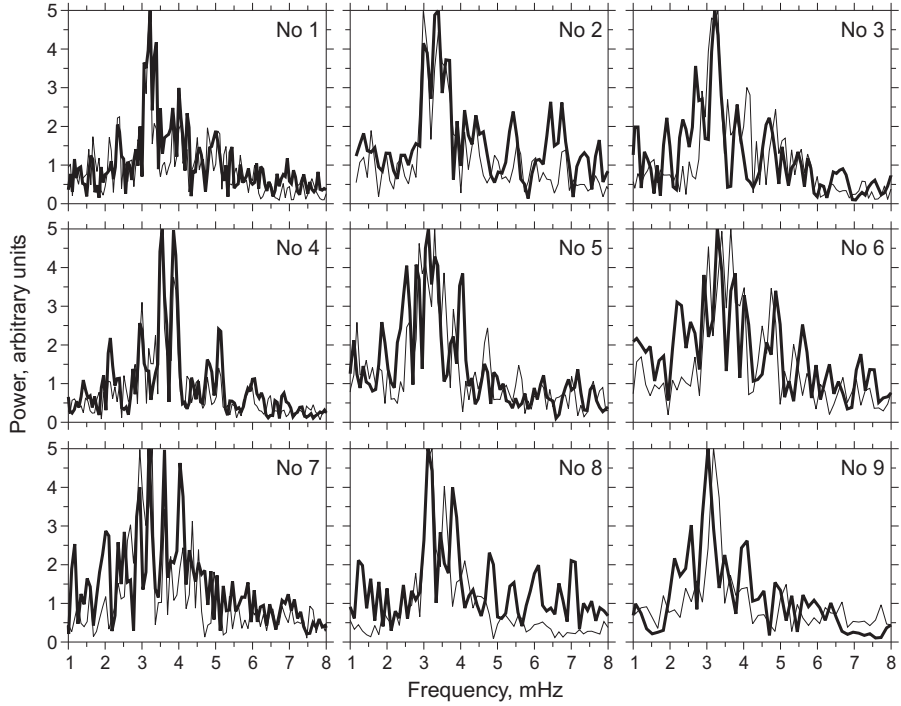}}
\caption{Power spectra of LOS velocity, observed in He\,{\sc i} (thick line) and Si\,{\sc i} (thin line),
 and normalized to their maximum values.}
  \label{F-PowerSpectra}
\end{figure}

\begin{figure}
 \centerline{\includegraphics[clip=]{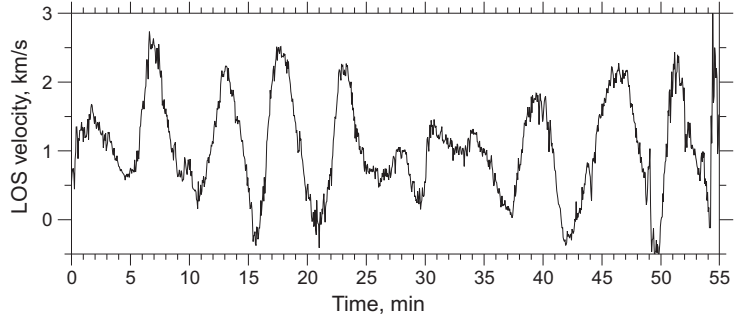}}
\caption{Unfiltered signal of chromospheric LOS velocity in He\,{\sc i} (facula 9).}
  \label{F-NonFiltered}
\end{figure}

In this article, we have the opportunity to compare results of observations, obtained in the same spectral
lines (He\,{\sc i} 10\,830\,\AA{} and Si\,{\sc i} 10\,827\,\AA{}). Even preliminary analysis revealed that power spectra of the LOS-velocity
oscillations looked very similar (Figure~\ref{F-PowerSpectra}). This similarity implies that the frequency of oscillations at the height of the
He\,{\sc i} 10\,830\,\AA{} line formation is identical to that at the height of the Si\,{\sc i} 10\,827\,\AA{} line formation. According to our observations in
the He\,{\sc i} 10\,830\,\AA{} line, five-minute oscillations are well detected in the chromosphere of facula. Figure~\ref{F-NonFiltered} showing unfiltered signal of
the LOS velocity confirms this.

Amplitudes of the photospheric and chromospheric signals should be equalized to determine phase delay between
them when they are filtered at the central frequency of 3.5 mHz. To do this we calculated the average amplitude of
signals $\mathrm{\textit{A}_{ph}}$ (for photospheric signals) and $\mathrm{\textit{A}_{ch}}$ (for chromospheric signals) for the entire time series.
The amplitude of the photospheric LOS-velocity signal was multiplied by  $\mathrm{\textit{K}=\textit{A}_{ch}/\textit{A}_{ph}}$, and both signals were then superimposed in the common figure;
 this allowed us to estimate roughly the time lag between the signals. Values of the amplitude multiplication factor of the photospheric LOS-velocity $[K]$ for different faculae are in
the interval from 1.2 to 8 and are presented in Table~\ref{T-TimeSeries}. One could suppose that
this spread of $K$ is not random and is probably connected with a feature of magnetic-field structure (magnetic-field strength, filling factor, inclination to vertical, magnetic shadow, and the presence of an admixture of opposite polarity).
Apparently more attention should be paid to this problem in future.

\begin{figure}
 \centerline{\includegraphics[clip=]{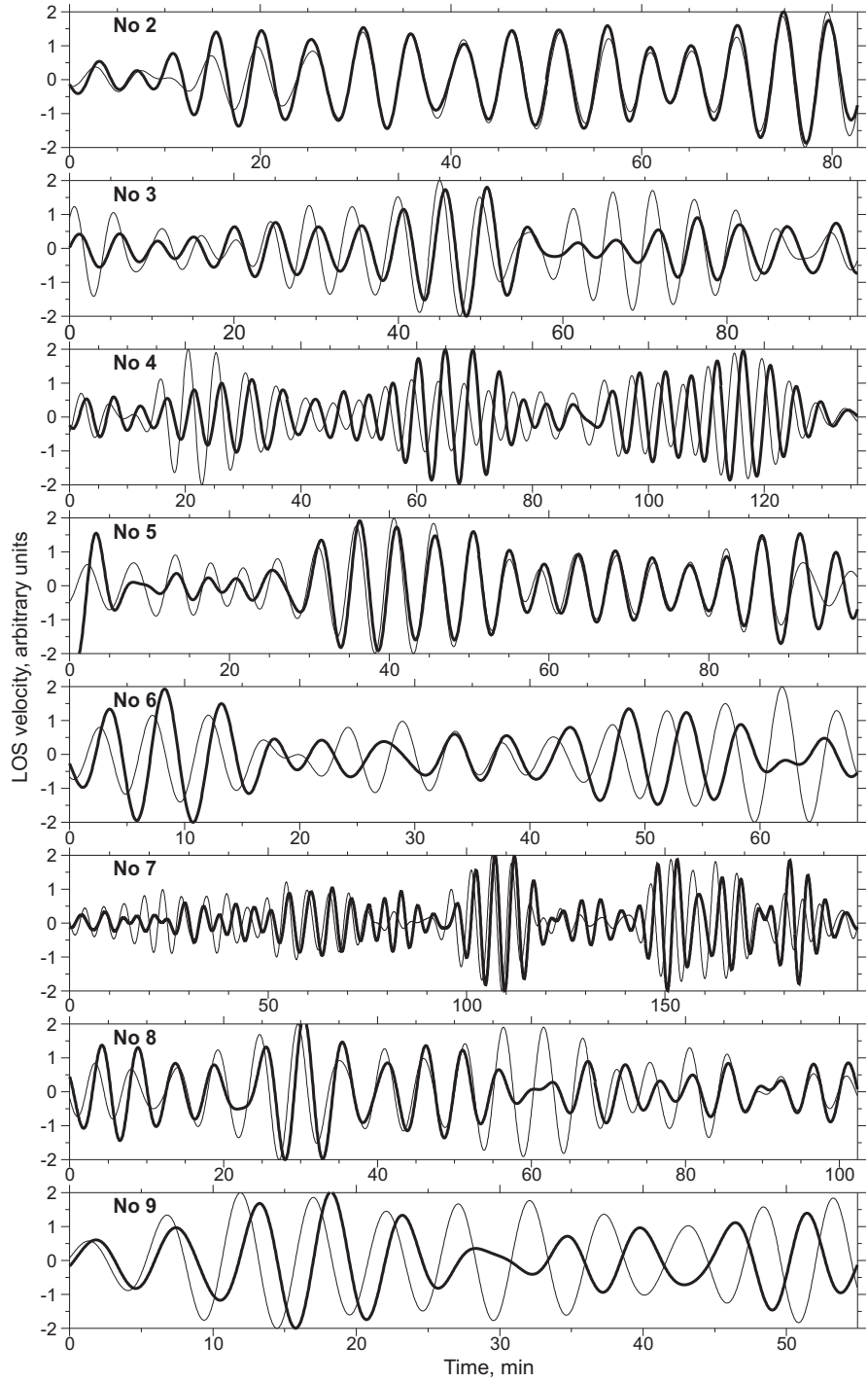}}
\caption{Phase relation between photospheric (thin line) and chromospheric (thick line) {unshifted} signals
of LOS velocity in different faculae. The signals have been filtered in a 1 mHz band around 3.5 mHz. Panel numbers are in accordance with Table\,\ref{T-TimeSeries}.}.
\label{F-PhaseRelation}
\end{figure}

\begin{figure}
 \centerline{\includegraphics[clip=]{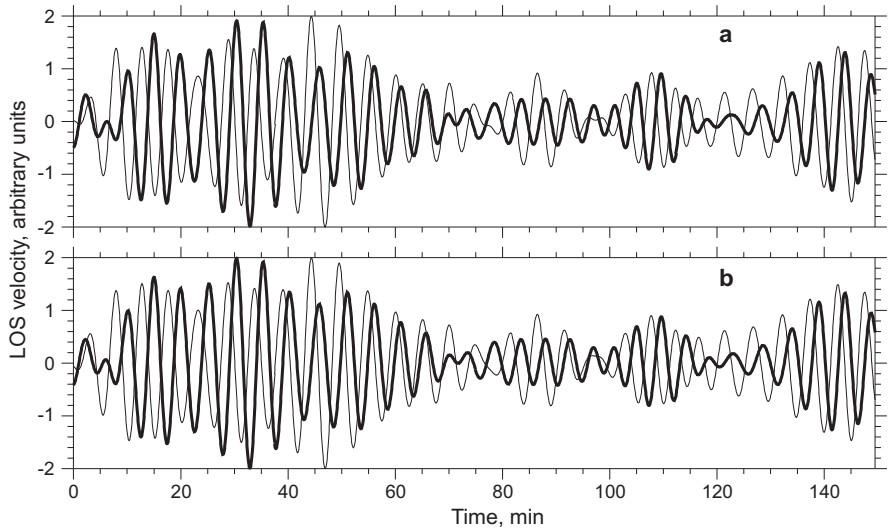}}
\caption{Examples of LOS-velocity signals (facula 1), filtered in different frequency bands around 3.5 mHz.
a)1 mHz band; b)0.2 mHz band.}
  \label{F-SignalsExemple}
\end{figure}

To determine the lag more accurately, we shifted the photospheric signal to later times in step of five seconds up to a maximum of 400 seconds and found the cross-correlation maximum that corresponded to the average time lag $[\delta{t}]$ in the series. Table\,\ref{T-TimeSeries} shows values $[\delta{t}]$ for
each analysed time series. As would be expected, the chromospheric signal of five-minute oscillations of the LOS velocity
lags behind the photospheric in the majority of faculae under investigation. In Figures~\ref{F-PhaseRelation} and \ref{F-SignalsExemple}, graphs of unshifted signals of LOS velocity are given. The time lag differs for different
series and sometimes can vary within one time series (Figure~\ref{F-PhaseRelation}). However, it never reaches values presented
in \inlinecite{Centeno09}. According to our measurements, the average lag is about 50 seconds. In Table~\ref{T-TimeSeries}, the spread of lag values obtained for different parts of each facula is given. The spread depicts the maximum and minimum of the lag values calculated for several spatial elements of the facula. Only those elements were taken into account in which five-minute oscillations are well detected in both the photosphere and the chromosphere. It is worthy of note that the wavetrain structure of LOS velocity excludes possible error in a value divisible by period when making quantitative estimation
of the lag (\textit{e.g.} see Figure~\ref{F-PhaseRelation}). We also tried to narrow the frequency filtering band of five-minute oscillations
from 1 mHz down to 0.2 mHz. No significant difference was observed afterwards, and the lag did not change (Figure~\ref{F-SignalsExemple}).

\begin{figure}
 \centerline{\includegraphics[clip=]{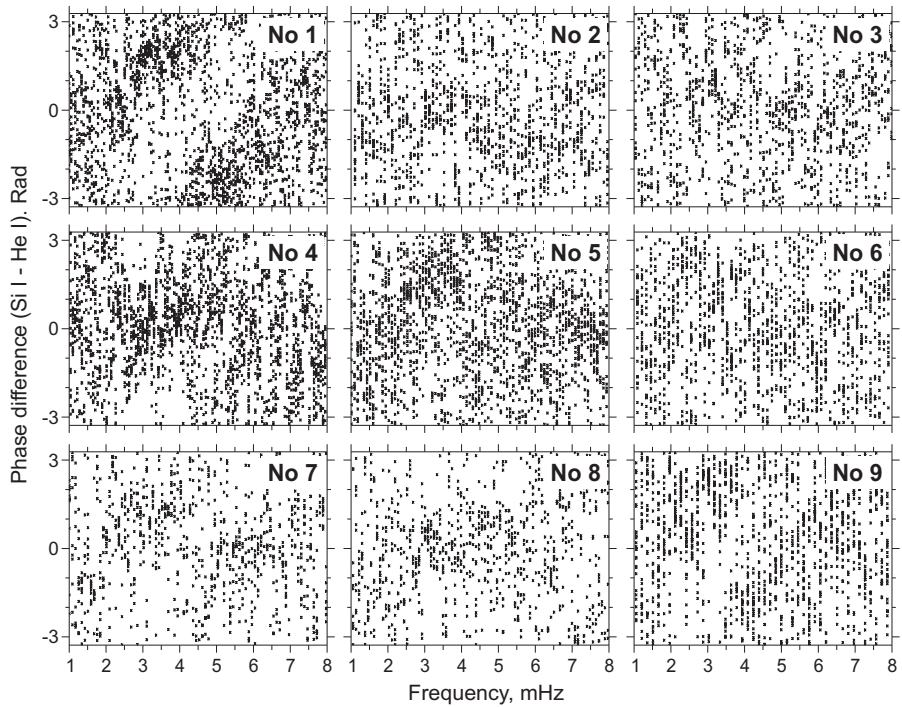}}
\caption{Phase difference between photospheric and chromospheric signals of LOS velocity  as a function of frequency. Each mark means the phase difference of LOS-velocity oscillations (He\,{\sc i}\,--\,Si\,{\sc i}) for each facula 2$''$ element along the entrance slit and particular frequency. Panel numbers are in accordance with Table\,\ref{T-TimeSeries}.}
  \label{F-PhaseDifference}
\end{figure}

We calculated the phase-difference spectra for each facula studied
(Figure~\ref{F-PhaseDifference}). Each mark on the Figure~\ref{F-PhaseDifference} means the phase difference of LOS-velocity oscillations (He\,{\sc i}\,--\,Si\,{\sc i}) for each facula 2$''$ element along the entrance slit and particular frequency. Each phase-difference spectrum was plotted over the whole facula (from 10 to 30 points). As was to be expected, phase-difference spectra seem ambiguous as a result of the smallness of the average lag measured. Only in four spectra (1,4,7,9) weak indications of propagating oscillations can be seen. We do not exclude the possibility that our measurements of the photospheric LOS velocity contains a contribution of small non-magnetic inclusions located along the entrance aperture.

Thus, our investigation of facular oscillations in the He\,{\sc i} 10\,830\,\AA{} and Si\,{\sc i} 10\,827\,\AA{} lines enables two propositions
to be made. The first proposition: we are in full agreement with \inlinecite{Khomenko08} and \inlinecite{Centeno09} about the fact
that spectra of the LOS-velocity oscillations observed in the He\,{\sc i} and Si\,{\sc i} lines look very similar.
The second proposition: in contradiction with the results obtained by these authors: the temporal lag of the chromospheric
signal does not exceed a period and is about 50 seconds on average.

The extreme similarity between spectra can be interpreted as a result of the fact that all oscillations from
the upper photosphere successfully reach the upper chromosphere.
At the same time, this similarity can be caused by closer heights of the He\,{\sc i} and Si\,{\sc i} line formation in facular regions.
Possible decrease in the height of the He\,{\sc i} 10\,830\,\AA{} line formation in faculae was noticed by \inlinecite{Livshits76}.
In \inlinecite{Kobanov11} we showed how similar oscillation spectra in faculae can be when the difference
in heights is only 300\,--\,400 km, using observations in the Ba\,{\sc ii} 4554\,\AA{} and Fe\,{\sc i} 4551\,\AA{} lines. Noteworthy is that the spatial
averaging over the entire facula in this case does not make the spectra dissimilar.

If the  velocity of the upward wave propagation is 4\,--\,6 $\mathrm {km\,s^{-1}}$ (as in \opencite{Centeno09}), the difference in heights of the
He\,{\sc i} and Si\,{\sc i} line formation in faculae will be about 250 km, considering our measured time lag of about 50 seconds.
At the same time, if the difference in formation heights of these lines is 1500 km (as in \opencite{Centeno09}), the velocity of the upward wave propagation will be about 30 $\mathrm {km\,s^{-1}}$. This velocity is six times greater than that mentioned
above.

Earlier the same problems occurred while researching oscillations observed at two levels of the undisturbed atmosphere. \inlinecite{Mein76} and \inlinecite{Mein77} explained a high phase velocity of about 30 $\mathrm {km\,s^{-1}}$ in the chromosphere by restoring forces in the presence of magnetic fields. \inlinecite{Fleck94}, analyzing observations in He\,{\sc i} 10\,830\,\AA{} and Mg\,{\sc i} 8807\,\AA{}, discussed the two alternatives that we cited above for faculae. In addition, \inlinecite{Fleck89} considered a variant in which it is supposed that the measured phase-difference forms in the range between temperature minimum and 1000 km. They suggested that the oscillations become standing at heights of more than 1000 km. \inlinecite{Gouttebroze99} found that variations of intensity in several UV spectral lines, observed with the SUMER onboard SOHO, demonstrated features of evanescent or standing waves.

If it is assumed that, within the cavity observed, both standing and propagating waves exist at the same time (quite an ordinary situation for any real resonator being the source of oscillations), then the phase lag being measured will depend on the contribution of each of the components. Their rate is determined by the degree of penetrability of the boundaries of the resonator that in real conditions changes following local changes in temperature, pressure and the magnetic field. Previously, \inlinecite{Kobanov11} indicated such a possibility when explaining the inadequacy of phase lags obtained from observations in H$\alpha$ and Fe\,{\sc i} 6569\,\AA{}.
Our preliminary measurements of phase difference between velocity and intensity oscillations showed that the lag of 90$^\circ$ is more characteristic of the signals measured in He\,{\sc i} 10\,830\,\AA{} than for those measured in Si\,{\sc i} 10\,827\,\AA{}. In the case of acoustic waves, this may imply that standing waves make a major contribution to oscillations observed at the level of He\,{\sc i} 10\,830\,\AA{} formation.

There is a contradiction that can be resolved with the supplementary information specifying both formation
depths of these lines and a possible propagation velocity of oscillations. In the future it is essential to obtain useful information on the structure and dynamics of the magnetic field in the faculae being studied. It would be useful to investigate the differences in spectral oscillations of faculae in active regions (AR) and faculae outside AR (for example, in polar regions) in more detail. It is desirable also to carry out a cycle of observations of limb faculae simultaneously with spectroscopic and filter methods. In any case, one should consider the results of our analysis carried out for nine faculae.

\section{Conclusion} 
      \label{S-Conclusion}
We carried out an investigation of features of periodic oscillations in nine facula regions for two heights
(He\,{\sc i} 10\,830\,\AA{} and Si\,{\sc i} 10\,827\,\AA{}). Power spectra of photospheric and chromospheric oscillations of the LOS velocity
were found to be very similar. According to our measurements, the lag of the chromospheric signal relative to
the photospheric one varies from -12 to 100 seconds and is, on average,  about 50 seconds for five-minute oscillations.

The features that  we have revealed can be interpreted in two ways: either a difference in heights of the He\,{\sc i} and Si\,{\sc i}
lines formation in faculae that is substantially smaller than the generally accepted estimate (1500 km), or propagation
velocity of five-minute oscillations above faculae is several times greater than the commonly used value of 4\,--\,6 $\mathrm{km\,s^{-1}}$. Another possible explanation supposes that within the cavity being observed the standing and propagating waves exist at one and the same time. In this instance the phase lag measured will depend on the contribution of each of the components.

In any case, this raises new questions as to whether or not the existing models correspond to modern observations.

\acknowledgements The work is supported by RFBR grants 08-02-91860-RS-a and 10-02-00153-a, the
Federal Agency for Science and Innovation (State contract 02.740.11.0576) and Basic Research Program
No. 16 (part 3) of the Presidium of the Russian Academy of Sciences.


\bibliographystyle{spr-mp-sola}

\end{article}
\end{document}